\documentclass[prd,twocolumn,showpacs,aps,epsbox,groupedaddress,eqsecnum]{revtex4}
\usepackage{amsfonts,latexsym,eucal}
\usepackage{graphicx}
\usepackage{epsfig}
\usepackage{color}
\def\beq{\begin{equation}}
\def\eeq{\end{equation}}
\def\bea{\begin{eqnarray}}
\def\eea{\end{eqnarray}}
\def\nn{\nonumber\\}

\def\pa{\partial}
\def\tr{\tilde R}
\def\tt{\tilde\tau}
\def\ra{\rightarrow}

\begin{document}
\draft
\title{Gravastar Shadows}
\author{Nobuyuki Sakai}
\email{nsakai@yamaguchi-u.ac.jp}
\affiliation{Faculty of Science, Yamaguchi University, Yamaguchi 753-8512, Japan}
\author{Hiromi Saida}
\email{saida@daido-it.ac.jp}
\affiliation{Department of Physics, Daido University, Minami-ku, Nagoya 457-8530, Japan}
\author{Takashi Tamaki}
\email{tamaki@ge.ce.nihon-u.ac.jp}
\affiliation{Department of Physics, General Education, College of Engineering, 
Nihon University, Tokusada, Tamura, Koriyama, Fukushima 963-8642, Japan}

\begin{abstract}
Direct observation of black holes is one of the grand challenges in astronomy.
If there are super-compact objects which possess unstable circular orbits of photons, however, it may be difficult to distinguish them from black holes by observing photons.
As a model of super-compact objects, we consider a gravastar (gravitational-vacuum-star) which was originally proposed by Mazur and Mottola.
For definiteness, we adopt a spherical thin-shell model of a gravastar developed by Visser and Wiltshire, which connects interior de-Sitter geometry and exterior Schwarzschild geometry.
We find that unstable circular orbits of photons can appear around the gravastar.
Then, we investigate the optical images of the gravastar possessing unstable circular orbits, with assuming the optically transparent surface of it and two types of optical sources behind the gravastar: (i) an infinite optical plane and (ii) a companion star.
The main feature of the image of (i) is that a bright disk and a dark thick ring surrounding the disk appear in the center of the region which would be completely dark if the compact object was not the gravastar but Schwarzschild black hole.
Also in the case (ii), a small disk and arcs around the disk appear in the region which would be completely dark for the lensing image by Schwarzschild black hole.
Because characteristic images appear inside the gravastar in both cases, we could tell the difference between a black hole and a gravastar with high-resolution VLBI observations near future.
\end{abstract}

\pacs{04.40.Dg, 97.60.Lf, 97.60.Lf}
\maketitle

\section{Introduction}

Direct observation of black holes is one of the grand challenges in astronomy and will be achieved by VLBI observations in the near future \cite{VLBI}.
In such an observation, we expect to observe the images of optical/radio sources around a black hole, which is called black-hole shadows.
In black-hole spacetimes such as Schwarzschild spacetime, there exist unstable circular orbits of photons, which play an important role generating shadows.
First, if optical sources are extended behind a black hole, one cannot detect photons which have passed inside the unstable circular orbits because any photon passing the unstable circular orbits inward eventually falls into the event horizon.
Second, because null geodesics wind several times in the vicinity of unstable circular orbits, one could observe brightening there when gas falls into the black hole.
Therefore, the direct observation of black holes usually means observation of geometry in the vicinity of unstable circular orbits of photons.

These natures naturally give rise to a question: are there any super-compact objects which possess unstable circular orbits of photons?
If such super-compact objects exist, one cannot claim to have seen a black hole even if one observes temporal brightening of gas falling into the object.
Therefore, it is important to study the possibility of super-compact objects and their observational consequences.

As a model of super-compact objects, we consider a model of gravastars.
Gravastars were originally proposed by Mazur and Mottola \cite{MM} as a new final state of gravitational collapse of stars, that is, an alternative to black holes. In this model an interior de Sitter region and an exterior Schwarzschild background are connected by a shell of stiff matter ($p=\rho$). Although their formation process is unclear, the idea is fascinating because it could solve two fundamental problems of black holes:  singularity problem and information loss paradox.

As we shall show in Sec.\ II, we find that some gravastar solutions possess unstable circular orbits of photons. 
This result indicates that it is difficult to distinguish those gravastars from black holes.
{Chirenti and Rezzolla \cite{CR} considered a question of how to tell a gravastar from a black hole.
They studied axial-perturbations on gravastars and found that their quasi-normal modes of gravitational waves differ from those of black holes.
Broderick and Narayan \cite{BN} argued that, if observed black hole candidates with matter accretion were gravastars, they should heat up and emit radiation. With this thermal process they discussed observational constraints on gravastar models.

In this paper we tackle the same question in a different approach: can we tell a gravastar from a black hole by electromagnetic observations instead of gravitational wave observations?
Obviously the answer depends on the state of the surface: there are three possibilities.
\begin{itemize}
\item The surface emits electromagnetic waves. In this case one could detect the electromagnetic waves, and hence a gravastar can be distinguished from a black hole observationally.
\item The surface is black and does not emit electromagnetic waves. In this case there is no chance to identify a gravastar.
\item The surface is electromagnetically transparent.
\end{itemize}
Observational consequences in the last case is not clear and potentially important. Therefore, we investigate observational images of the gravastar under the assumption that the surface is electromagnetically transparent.

This paper is organized as follows.
In Sec. II, we reanalyze the thin-shell model of a gravastar and search stable solutions systematically.
In Sec. III, we derive null geodesic equations and find solutions which possess unstable circular orbits of photons.
In Sec. IV, we solve the null geodesic equations numerically to obtain the images of optical sources behind the gravastar.
Section V is devoted to concluding remarks.

\section{Thin-shell model }

We consider the spherical thin-shell model of a gravastar developed by Visser and Wiltshire \cite{VW}.
Here we adopt their model.
The inside  is a part of de Sitter spacetime,
\bea\label{dS}
&&ds^2=-A_-dt_-^2+{dr_-^2\over A_-}+r_-^2(d\theta^2+\sin^2\theta d\varphi^2),\nn
&&{\rm with}~~~A_-(r_-)\equiv1-H^2r_-^2,
\eea
and the outside is a part of Schwarzschild spacetime,
\bea\label{Sch}
&&ds^2=-A_+dt_+^2+{dr_+^2\over A_+}+r_+^2(d\theta^2+\sin^2\theta d\varphi^2),\nn
&&{\rm with}~~~A_+(r_+)\equiv1-{r_g\over r_+},
\eea
where $r_g$ is a gravitational radius.
We have denoted field variables on the outside (inside) by superscripts or subscripts $+~(-)$.
To describe the geometry in the vicinity of the boundary hypersurface $\Sigma$, we introduce a Gaussian normal coordinate system,
\bea\label{GNC}
ds^2&=&dn^2+\gamma_{ij}^\pm x^idx^j\nn
&=&dn^2-\alpha_\pm(n,\tau)^2d\tau^2\nn
&&+{r_\pm}(n,\tau)^2(d\theta^2+\sin^2\theta d\varphi^2),
\eea
in which $n=0$ corresponds to $\Sigma$. $\alpha$ is normalized by $\alpha_\pm(0,\tau)=1$ so that $\tau$ implies the proper time of $\Sigma$. $R(\tau)\equiv{r_\pm}(0,\tau)$ denotes the areal radius of $\Sigma$.
We suppose that $\Sigma$ contains infinitesimally thin matter,
\beq
S^i_j\equiv\int^{+0}_{-0}T^i_jdn={\rm diag}(-\sigma,~\varpi,~\varpi),
\eeq
where $\sigma$ and $\varpi$ are the surface energy density and the surface pressure, respectively.
Following Visser and Wiltshire \cite{VW}, we assume 2+1 dimensional stiff matter:
\beq\label{stiff}
\varpi=\sigma.
\eeq

What we are looking for is static and stable solutions of a gravastar. As for their stability, we only consider stability against spherical perturbations.
Stability against non-spherical perturbations is important as well, but it is beyond the present work.
In the following we derive the equations of motion of the shell to find static and stable solutions.

Following the Israel's formalism \cite{Israel,BGG}, we can obtain the junction conditions at $\Sigma$ as follows.
The metric continuity $\gamma_{ij}^+=\gamma_{ij}^-$ implies
\bea\label{continuity}
&&R=r_+=r_-,\\
&&d\tau^2=A_+dt_+^2-{dr_+^2\over A_+}=A_-dt_-^2-{dr_-^2\over A_-}.
\eea
{The other junction conditions are reduced to the two equations. One is
\beq\label{jc}
\beta_--\beta_+=4\pi G\sigma R,
\eeq
where
\beq
\beta_\pm\equiv{\pa r_\pm\over\pa n}=\varepsilon_{\pm}\sqrt{\left({dR\over d\tau}\right)^2+A_{\pm}},~~
\varepsilon_\pm\equiv{\rm sign}{\pa r_\pm\over\pa n}.
\label{2ec}\eeq
In a spacetime without Schwarzschild horizon nor de Sitter horizon, $\varepsilon_\pm=+1$ because the areal radius $r_\pm$ always increases as $n$ increases. The other equation is}
\beq
{d\over d\tau}(\sigma R^2)+\varpi{d\over d\tau}(R^2)=0.
\eeq
For stiff matter (\ref{stiff}), we find
\beq
\sigma R^4={\rm const.}
\eeq

Introducing dimensionless quantities, 
\bea\label{rescale}
&&\tr\equiv{R\over r_g},~~~ \tt\equiv{\tau\over r_g},\nn
&&h\equiv r_gH,~~~
s\equiv{4\pi G\sigma R^4\over r_g^3}={\rm const.},
\eea
we rewrite (\ref{jc}) as
\beq\label{eom}
\left({d\tr\over d\tt}\right)^2+U(\tr)=0,
\eeq\beq\label{U}
U(\tr)\equiv1-{h^2\tr^2\over2}-{1\over2\tr}-{s^2\over4\tr^6}
-{\tr^6\over4s^2}\left(h^2\tr^2-{1\over\tr}\right)^2.
\eeq

\begin{figure}
\includegraphics[scale=.45]{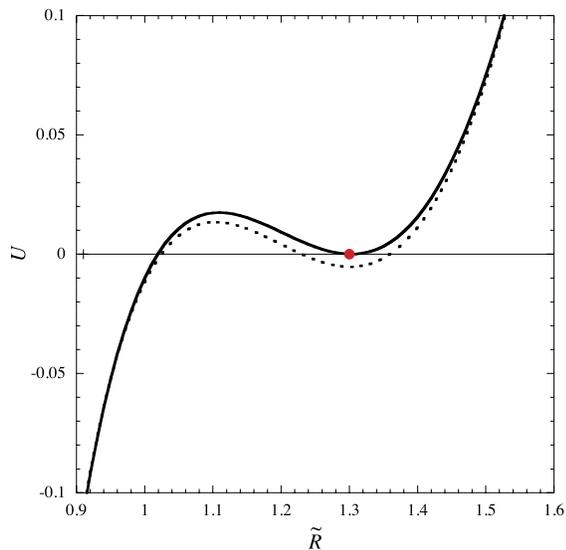}
\caption{Graph of the potential function $U(\tr)$. The red dot on the bold line represents a static solution, where we put $h=0.4$ and $s=0.8212081$. If we give perturbations on the static solution by changing $r_g\ra 1.003r_g$, we obtain the potential indicated by the dotted line. The perturbed potential allows an oscillating solution, which implies that the original static solution is stable.}
\end{figure}

\begin{figure}
\includegraphics[scale=.45]{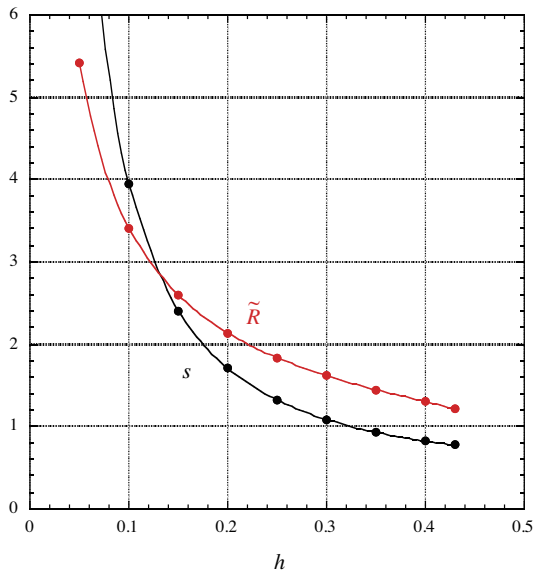}
\caption{Parameters $s$ and $\tr$ as a function of $h$, when $U(\tr)$ has a local minimum on $U(\tr)=0$ as shown in Fig. 1.
Only these parameter values allow for static and stable gravastar solutions.
$h$ has an upper limit, $h_{\rm max}\approx0.43$.}
\end{figure}

Figure 1 shows an example of the potential function $U(\tr)$ which allows a static and stable solution. The red dot on the bold line represents a static solution. If we give perturbations on the static solution by changing $r_g\ra 1.003r_g$, we obtain the potential indicated by the dotted line. The perturbed potential allows an oscillating solution, which implies that the original static solution is stable against perturbations with increasing mass.

As Fig. 1 indicates, static and stable solutions exist if and only if $U(\tr)$ has a local minimum on $U(\tr)=0$.
This condition gives a constraint on the three parameters $h,~s$ and $\tr$.
We survey static and stable solutions and show their parameters $s$ and $\tr$ as a function of $h$ in Fig. 2.
$h$ has an upper limit, $h_{\rm max}\approx0.43$, because the local minimum of $U(\tr)$ disappears or cannot be located on $U(\tr)=0$ when $h>h_{\rm max}$.

\section{Null Geodesic Equations}

In this section we derive null geodesic equations with boundary conditions for the static gravastar spacetime obtained in \S II.
We denote the affine parameter and the null vector by $\lambda$ and $k^\mu=dx^\mu/d\lambda$, respectively; then the geodesic equations are generally expressed as
\beq\label{geodesic}
{dk^\mu\over d\lambda}+\Gamma^\mu_{\nu\rho}k^\nu k^\rho=0,~~~
{\rm with}~~~k_\mu k^\mu=0.
\eeq
The geodesics in the $\theta=\pi/2$ plane for the outside $(+)$ and the inside $(-)$ are given by
\beq\label{geo1}{
{d\over d\lambda_\pm}(A_\pm k^t_\pm)=0,~~~
{d\over d\lambda_\pm}(r_\pm^2k^\varphi_\pm)=0,}
\eeq\beq
{1\over\sqrt{A_\pm}}{d\over d\lambda_\pm}\left({k_\pm^r\over\sqrt{A_\pm}}\right)+{dA_\pm\over dr_\pm}{(k^t_\pm)^2\over2}-r_\pm(k^\varphi_\pm)^2=0,
\label{geo2}\eeq
\beq
-A_\pm(k^t_\pm)^2+{(k^r_\pm)^2\over A_\pm}+r_\pm^2(k^\varphi_\pm)^2=0.
\label{geo3}\eeq
Because Eq.(\ref{geo2}) is also derived by (\ref{geo1}) and (\ref{geo3}), we do not have to solve it.
Equations (\ref{geo1}) are integrated as
\beq\label{geo11}
A_\pm k^t_\pm={\rm const}.\equiv E_\pm,~~~
r_\pm^2k^\varphi_\pm={\rm const}.\equiv L_\pm,
\eeq
and then (\ref{geo3}) becomes
\beq\label{geo21}
(k_\pm^r)^2+{A_\pm L_\pm^2\over r_\pm^2}=E_\pm^2.
\label{geo31}\eeq
It follows from (\ref{geo11}) and (\ref{geo21}) that
\beq
{dr_\pm\over d\varphi}={k^r_\pm\over k^\varphi_\pm}={r_\pm^2k_\pm^r\over L_\pm}
=\pm r_\pm\sqrt{\left({E_\pm r_\pm\over L_\pm}\right)^2-A_\pm},
\label{geo42}\eeq
which gives null geodesics in the exterior and interior regions of the gravastar.

\begin{figure}
\includegraphics[scale=.5]{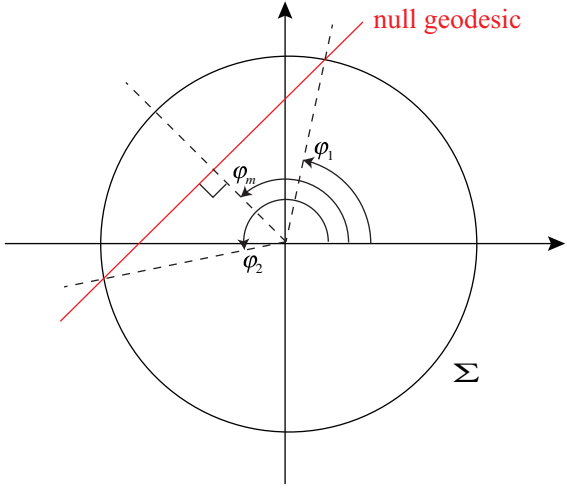}
\caption{Null geodesics penetrating the gravastar. $\varphi_1$ and $\varphi_2$ ($\varphi_1<\varphi_2$) denote the two crossing points. The integral constant $\varphi_m$ corresponds to the closest point to the center.}
\end{figure}

The equation for the interior region in (\ref{geo42}) is integrated as
\beq
r_-=r_m\sec(\varphi-\varphi_m),~~~
r_m\equiv\left({E_-^2\over L_-^2}+H^2\right)^{-\frac12},
\eeq
where $\varphi_m$ is an integral constant.
Since a black hole horizon does not exist and the surface of the gravastar is transparent in our situation, any incident light ray to the gravastar has to penetrate the gravastar, as shown in Fig.\ 3.
Therefore, there are two crossing points of the penetrating null geodesic with the surface of gravastar $\Sigma$.
Let $\varphi_1$ and $\varphi_2~(\varphi_1 < \varphi_2$) denote the $\varphi$-coordinate values of those two crossing points, they are determined by
\beq
\varphi_m=\varphi_1+\arccos{r_m\over R}=\varphi_2-\arccos{r_m\over R}.
\eeq
On the other hand, the equation for the exterior region in (\ref{geo42}) cannot be integrated analytically.
However, the asymptotic solution at $r\ra\infty$ is obtained by putting $A_+\ra1$:
\beq\label{asysol}
{r_+={L_+\over E_+}}\sec(\varphi-\varphi_c),
\eeq
where $\varphi_c$ is an integral constant.

Next, we discuss the boundary conditions of $k^\mu$ at $\Sigma$. 
In the case of a static gravastar, $R=$const., the relation between the Gaussian normal coordinates (\ref{GNC}) and the outer/inner coordinates (\ref{dS}) and (\ref{Sch}) are given by
\bea
d\tau^2&=&A_+dt_+^2=A_-dt_-^2,\nn
R^2d\varphi^2&=&r_+^2d\varphi^2=r_-^2d\varphi^2.
\eea
Then we find
\beq\label{kbc1}
\sqrt{A_+}k^t_+=\sqrt{A_-}k^t_-,~~~k^\varphi_+=k^\varphi_-.
\eeq
With the help of the null condition (\ref{geo3}), we also obtain
\beq\label{kbc2}
{k^r_+\over\sqrt{A_+}}={k^r_-\over\sqrt{A_-}}.
\eeq
The relations among the integration constants are given by (\ref{continuity}), (\ref{geo11}) and (\ref{kbc1}),
\beq\label{kbc3}
L_+=L_-,~~~{E_+\over\sqrt{A_+}}={E_-\over\sqrt{A_-}}.
\eeq
Hereafter we denote $L_+$ and $L_-$ simply by $L$ because they are identical.

\begin{figure}
\includegraphics[scale=.45]{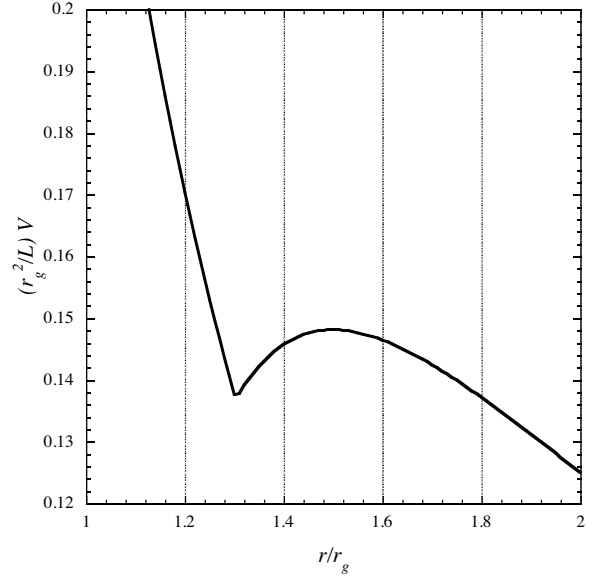}
\caption{Graph of the effective potential of null geodesics, $V(r)$,  for the static solution with $h=0.4~ (\tr=1.303$).
There are a stable circular orbit at $r=R$ and an unstable circular orbit at $r=1.5r_g$.}
\end{figure}

To make a qualitative discussion on photon trajectories, it is convenient to introduce the effective potential as follows.
Equation (\ref{geo21}) is rewritten as {
\beq\label{geo33}
\left({dr_\pm\over d\lambda_\pm}\right)^2+{L^2A_\pm\over r_\pm^2}=E_\pm^2.
\eeq
To discuss the dynamics with a continuous ``potential" by analogy with the Newtonian mechanics, we introduce unified variables as
\bea
r=r_-~~{\rm and}~&\lambda=\sqrt{{A_-(R)\over A_+(R)}}\lambda_-&~{\rm (inside)},\nn
r=r_+~~{\rm and}~&\lambda=\lambda_+&~{\rm (outside)},
\eea}
and define the effective potential as
\bea
V(r<R)&=&{L^2A_-\over r^2}{A_+(R)\over A_-(R)}=L^2{A_+(R)\over A_-(R)}\left({1\over r^2}-H^2\right),\nn
V(r>R)&=&{L^2A_+\over r^2}=L^2\left({1\over r^2}-{r_g\over r^3}\right).
\eea
Then we obtain the continuous equation of motion,
\beq
\left({dr\over d\lambda}\right)^2+V(r)=E_+^2.
\eeq

Figure 4 shows the effective potential for the static solution with $h=0.4$.
There are a stable circular orbit at $r=R$ and an unstable circular orbit at $r=1.5r_g$.
Contrary to the case of a black hole, even if a photon crosses the unstable circular orbit, $r=1.5r_g$, it eventually scattered away from it.

\section{Images of optical sources behind a gravastar}

\subsection{Basic features of null geodesics coming to observer}

\begin{figure}
\includegraphics[scale=.5]{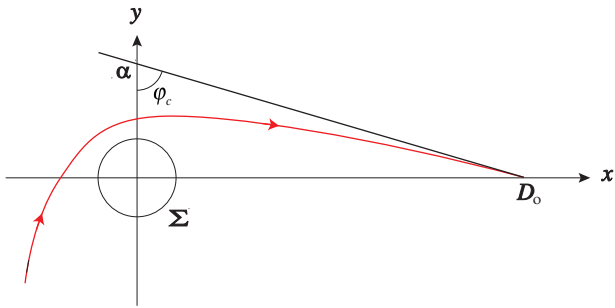}
\caption{The $z=0$ plane in the 3-space $(x,y,z)=(r\cos\varphi\sin\theta,r\sin\varphi\sin\theta,r\cos\theta)$.
The center of the gravastar is located at the origin and the observer at $(D_o,0,0)$. 
The intersection of the $y$-axis with the tangent to the ray at the observer is denoted by $(0,\alpha,0)$.
The integral constant $\varphi_c$ corresponds to the angle indicated by this figure: $\tan\varphi_c=D_o/\alpha$.
}
\end{figure}

In order to describe light trajectories, we define the rectangular coordinates $(x,y,z)=(r\cos\varphi\sin\theta,r\sin\varphi\sin\theta,r\cos\theta)$. We suppose that the center of the gravastar is located at the origin and the observer at $(D_o,0,0)~(\varphi=0)$. Figure 5 shows the $z=0~(\theta=\pi/2)$ plane.
If we make a coordinate rotation appropriately, any trajectory can be put on this plane.
On this plane, we denote the intersection of the $y$-axis with the tangent to the ray at the observer by $y=\alpha$.
In this rectangular coordinate system the asymptotic solution (\ref{asysol}) is rewritten as
\beq
x\cos\varphi_c+y\sin\varphi_c=\frac LE,
\eeq
where the $x$-intercept and the $y$-intercept are given by
\beq\label{Doalpha}
D_o=\frac LE\sec\varphi_c,~~~\alpha=\frac LE{\rm cosec}~\varphi_c,
\eeq
respectively. Recall that $\varphi_c$ is an integral constant defined by (\ref{asysol}).
Because (\ref{Doalpha}) indicates
\beq
\tan\varphi_c={D_o\over\alpha},
\eeq
we find that the integral constant $\varphi_c$ corresponds to the angle indicated by Fig.\ 5.

Furthermore, taking the limit of $D_o\ra\infty$, we obtain
\beq
\varphi_c\ra{\pi\over2},~~~\alpha\ra{L\over E}.
\eeq
Therefore, if $D_o$ is large enough, we can regard $L/E$ as the apparent length from the center as well as the impact parameter.

\begin{figure}
\includegraphics[scale=.47]{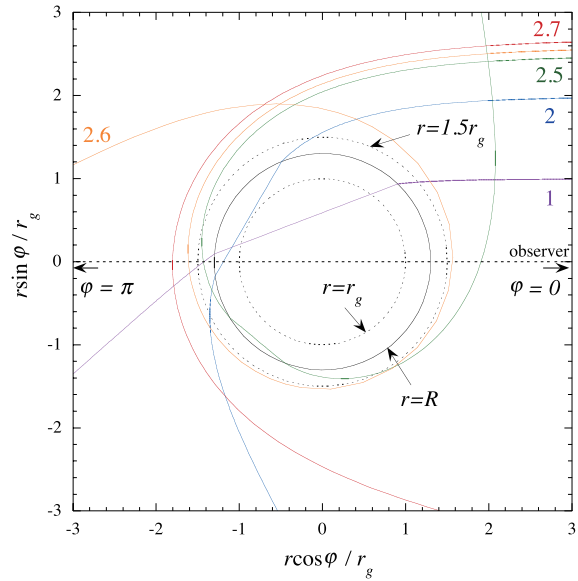}
\caption{Trajectories of photons which reach the observer for  the model with $h=0.4~(\tr=1.303$).
The observer is located at $r=1000r_g,~\varphi=0$, the right side of the figure.
We denote the positions of $r=r_g$ and $r=1.5r_g$ by dotted lines for reference.
We show five trajectories with $L/Er_g=1,~2,~2.5,~2.6$ and 2.7.}
\end{figure}

\begin{figure}
\includegraphics[scale=.47]{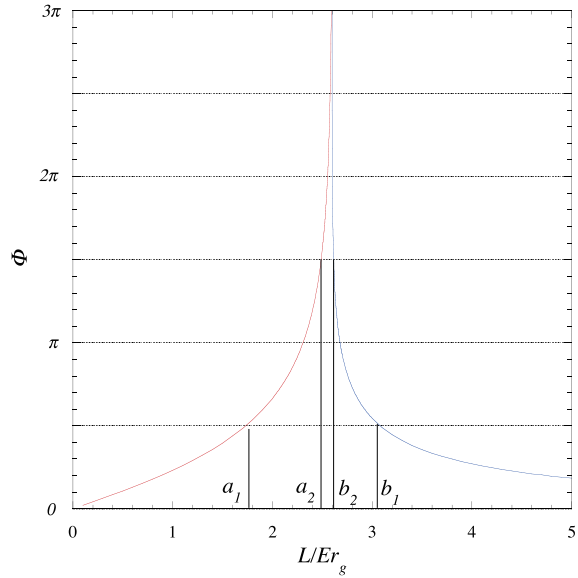}
\caption{Relation between the impact parameter $L/E$ and the deflection angle $\Phi$, which is defined by (\ref{Phi}), for the model with $h=0.4~(\tr=1.303$). The observer is located at $\varphi=0,~r=1000r_g$.
The peak corresponds to the geodesics which wind infinite times on the unstable circular orbit $r=1.5r_g$,
and diverges at $L/Er_g=1.5\sqrt{3}\approx2.5981$.
While the photons with $L/Er_g>1.5\sqrt{3}$ (blue line) travel only in Schwarzschild background,
the photons with $L/Er_g<1.5\sqrt{3}$ (red line) pass through the gravastar interior.
We denote the values of $L/Er_g$ which satisfy $\Phi=\pi(n-1/2)$ by $a_n,~b_n~(a_n<b_n)$.}
\end{figure}

In our numerical analysis below, we put an observer at $D_o=1000r_g$.
Figure 6 shows several trajectories of  photons which reach the observer for  the gravastar with $h=0.4$.

To understand more clearly how trajectories depend on  the impact parameter $L/E$, we define the deflection angle $\Phi$ by
\beq\label{Phi}
\Phi \equiv \varphi(r_g=1000 r_g)-\pi,
\eeq
which denotes the deflection measured from the opposite direction to $x$-axis.
This $\Phi$ is convenient for the next subsection.
Figure 7 shows the relation between $\Phi$ and $L/E$.
The peak corresponds to the geodesics which wind {infinite times on} the unstable circular orbit $r=1.5r_g$,
and diverges at $L/Er_g=1.5\sqrt{3}\approx2.5981$.
While the photons with $L/Er_g>1.5\sqrt{3}$ (blue line) travel only in Schwarzschild background,
the photons with $L/Er_g<1.5\sqrt{3}$ (red line) pass through the gravastar interior.
In pure Schwarzschild spacetime only geodesics with $L/Er_g>1.5\sqrt{3}$  (blue line) exist.
We also denote the values of $L/Er_g$ which satisfy
\beq
\Phi=\pi\left(n-\frac12\right),~~~(n=1,2,...).
\eeq
by $a_n,~b_n~(a_n<b_n)$.
For convenience we also define $a_0$ as $a_0=0$.

In the following we consider two types of optical sources behind the gravastar: an infinite optical plane and a companion.

\subsection{Image of an infinite optical plane behind a gravastar}

\begin{figure}
\includegraphics[scale=1.1]{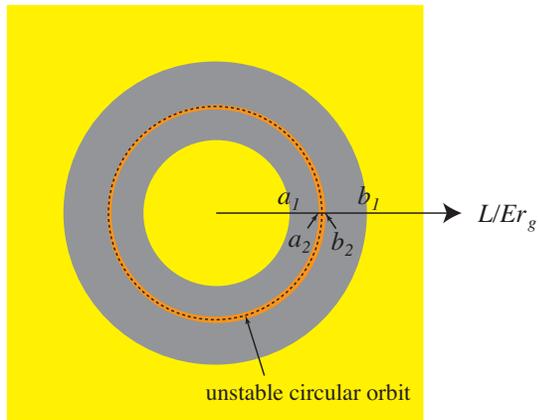}
\caption{Shadow cast by an infinite optical plane behind the gravastar with $h=0.4~(\tr=1.303)$.
This figure is depicted based on the numerical results in Fig.\ 7.
In the yellow domains the photons emitted by the infinite plane come to the observer.
We do not take account of light-dark contrast, which is generated by gravitational redshift of the photons.
In the dark domains $[a_1,a_2]$ and $[b_1,b_2]$, no photon comes to the observer from the optical infinite plane.
In the orange domain $[a_2,b_2]$  there are infinite numbers of bright and dark rings.
The dotted circle corresponds to the unstable circular orbit in Schwarzschild background.}
\end{figure}

An infinite plane is a simplified model of an extended gas behind the gravastar.
Because the deflection angle of photons which emit by the infinite plane and come to the observer should be in the range, 
\beq
a_{2n} <\Phi< a_{2n+1}~(n=0,1,2,...), 
\eeq
Figure 7 enables us to depict the shadow of the gravastar with $h=0.4$, as shown in Fig.\ 8.

The main feature is that a bright disk in $L/Er_g<a_1$ is surrounded by a dark domain $a_1<L/Er_g<a_2$.
In the dark domains $[a_1,a_2]$ and $[b_1,b_2]$, no photon comes to the observer from the optical infinite plane.
In the orange domain $[a_2,b_2]$  there are infinite numbers of bright and dark rings.
While the image outside the dotted circle is common to pure Schwarzschild spacetime, the image inside it is unique to the gravastar.
Here we do not take account of light-dark contrast, which is generated by gravitational redshift of the photons.

Because $a_1$ corresponds to the geodesics whose deflection angle is $\pi/2$,  the whole landscape behind the gravastar is reduced in the central yellow disk. Hence, we can interpret the gravastar as a concave lens with much reduction rate.

\begin{figure}
\includegraphics[scale=.47]{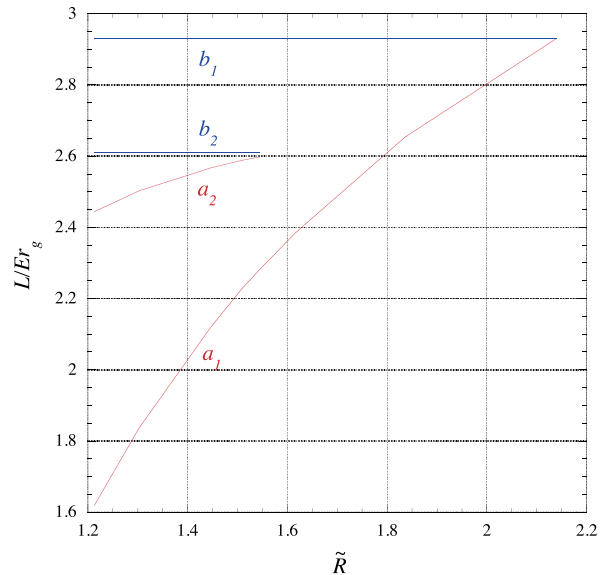}
\caption{Ring sizes of shadows cast by backlight for several models.
$b_1$ and $b_2$ are independent of or $\tr$ as long as they exist,  because they are determined by Schwarzschild geometry.
$a_1$ and $a_2$ are increasing functions of $\tr$, and they have upper limits, $a_2<b_2\approx2.61$ and $a_1<b_1\approx2.93$.
}
\end{figure}

Next, we discuss how the ring sizes depend on the model parameters.
Among the three parameters $h,~s$ and $\tr$, only one is independent, as shown in Fig.\ 2.
Therefore we show the ring sizes $a_1,~a_2,~b_2$ and $b_1$ as a function of $\tr$ in Fig.\ 9.
Here we extend our analysis to the case where there is no unstable circular orbit of photons, i.e., $\tr>1.5$.
$b_1$ and $b_2$ are independent of $h$ or $\tr$ as long as they exist,  because they are determined by Schwarzschild geometry.
$a_1$ and $a_2$ are increasing functions of $\tr$, and they have upper limits, $a_2<b_2\approx2.61$ and $a_1<b_1\approx2.93$.

There are merging points of $a_i$ and $b_i$, which we denote $\tr=\tr_i$.
We numerically find that $\tr_2\approx1.58$ and $\tr_1\approx2.14$.
The orange domain [$a_2,b_2$] in Fig.\ 8 disappears when $\tr>\tr_2$ and the dark domain [$a_1,b_1$] disappears when $\tr>\tr_1$.
The merging points are determined by the Schwarzschild geometry as follows.
For the photon trajectory of $b_i$, there is the closest point to the gravastar's center, whose position we denote by $r_+=$min$[r_+(b_i)]$.
The trajectory of $b_i$ cannot exist if the gravastar's radius is larger than min$[r_+(b_i)]$. We thus obtain the relation,
\beq
\tr_i={\rm min}\left[{r_+(b_i)\over r_g}\right].
\eeq

\subsection{Image by a companion star behind a gravastar}

\begin{figure}
\includegraphics[scale=.5]{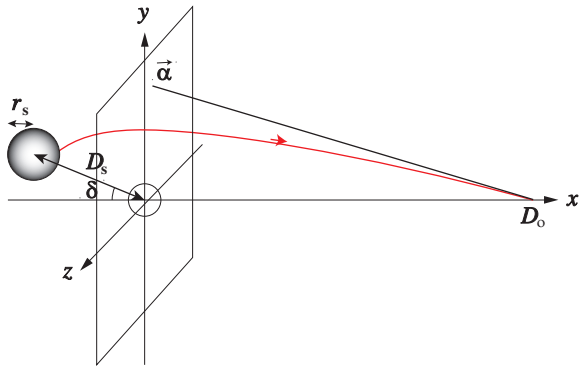}
\caption{Setting of our analysis of gravitational lens effects. We suppose a companion behind a gravastar.
We put the gravastar's center and the companion's center at the origin and on the $z=0$-plane, respectively.
We denote the distance between the companionfs center and the gravastar's center by $D_s$ and the radius of the companion by $r_s$.
The angle $\delta$ is defined as the angle between the direction of the companion's center and the opposite direction to $x$-axis.
The image $\vec\alpha=(\alpha_y,\alpha_z)$ is defined as the intersection of the $x=0$ plane with the tangent to the ray at the observer.}
\end{figure}

\begin{figure}
\includegraphics[scale=.36]{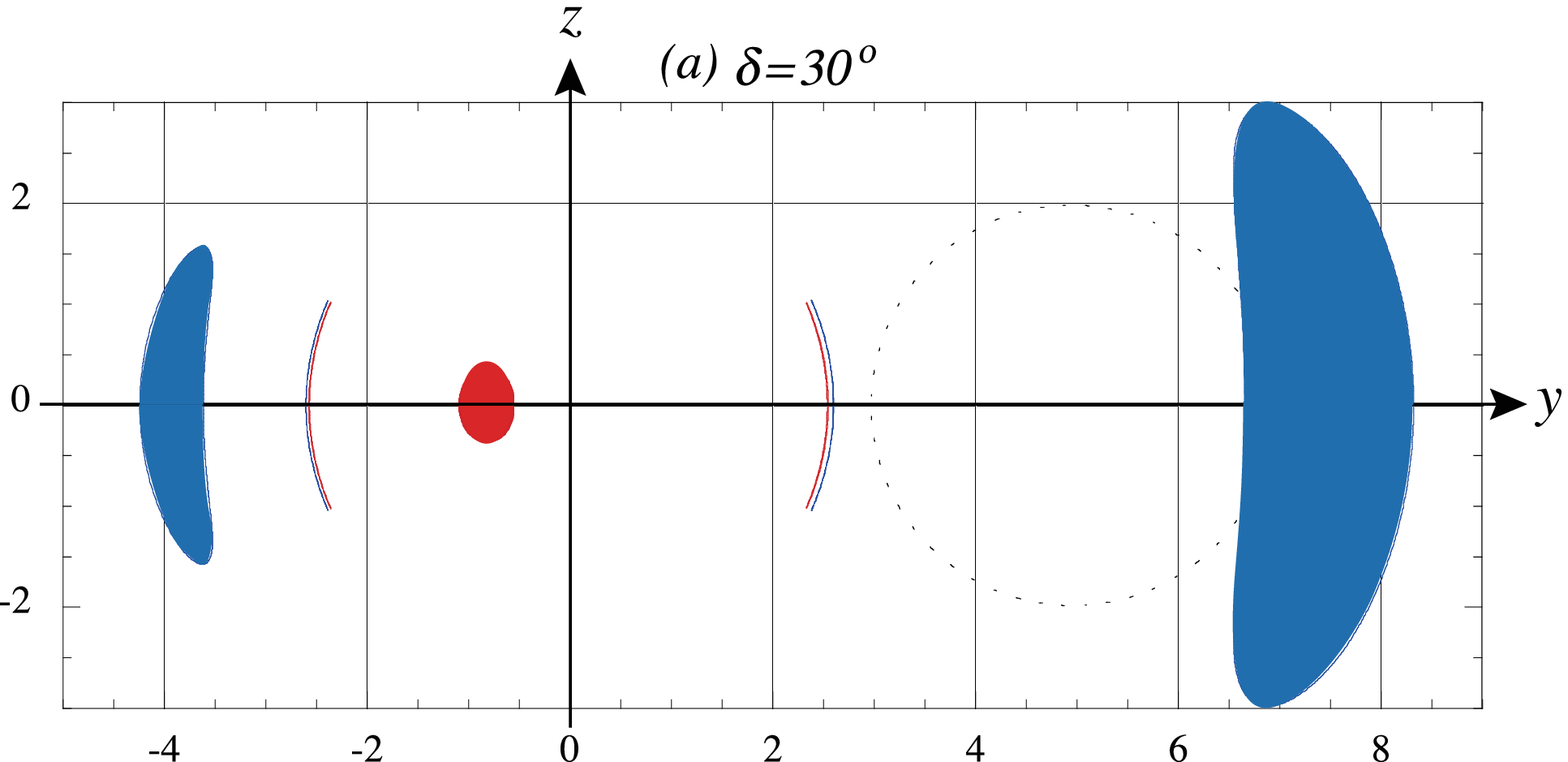}
\includegraphics[scale=.36]{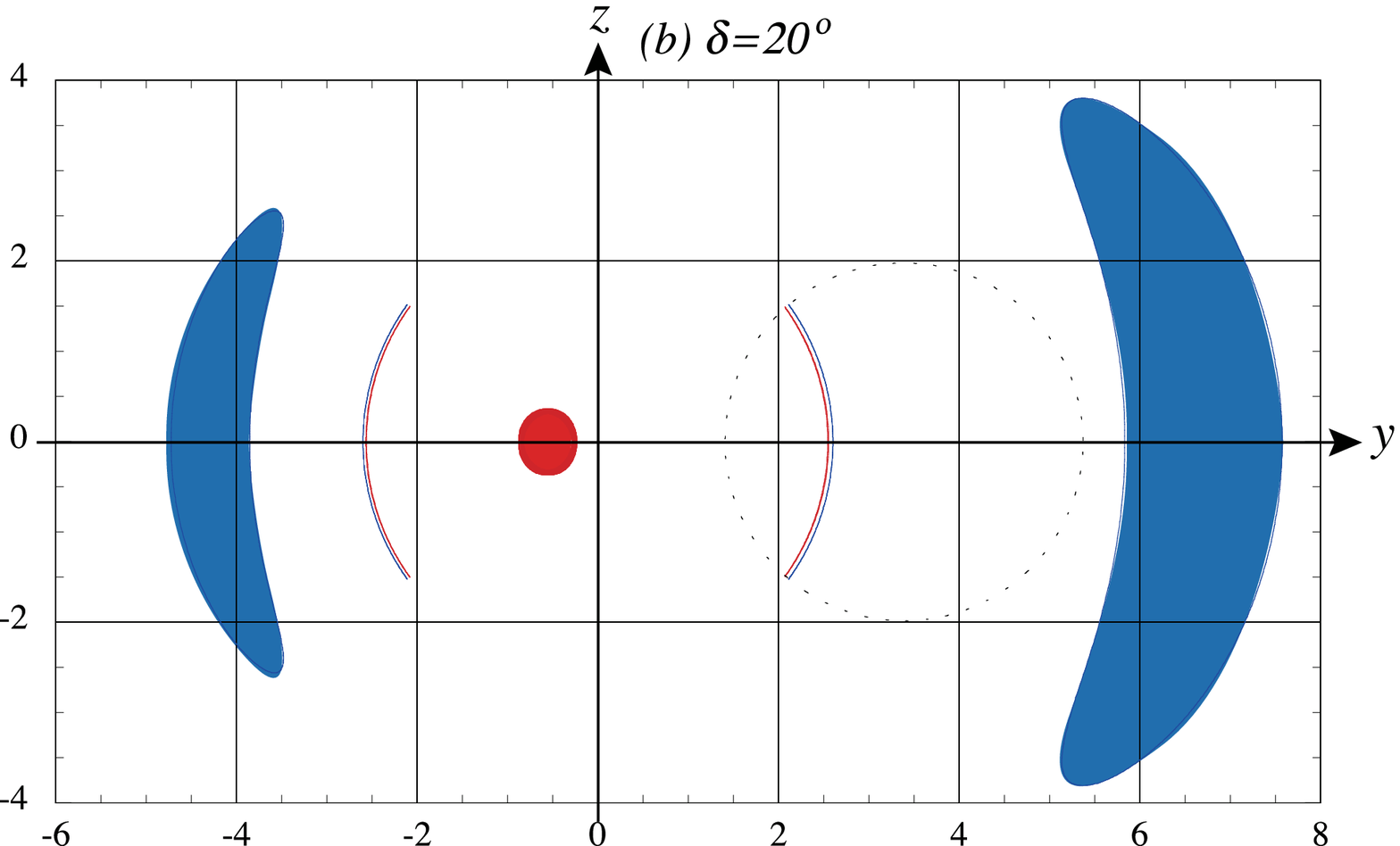}
\includegraphics[scale=.36]{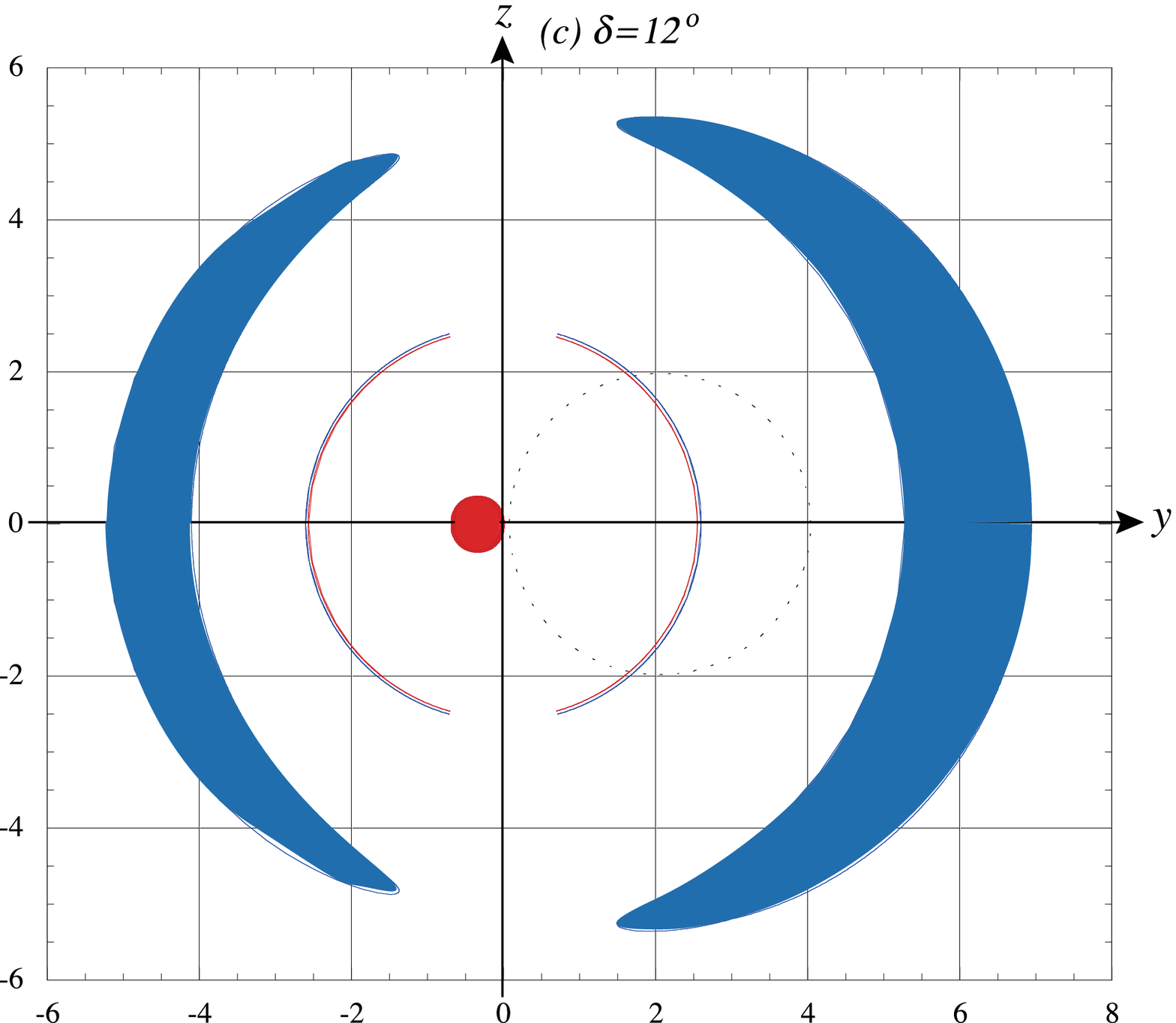}
\includegraphics[scale=.36]{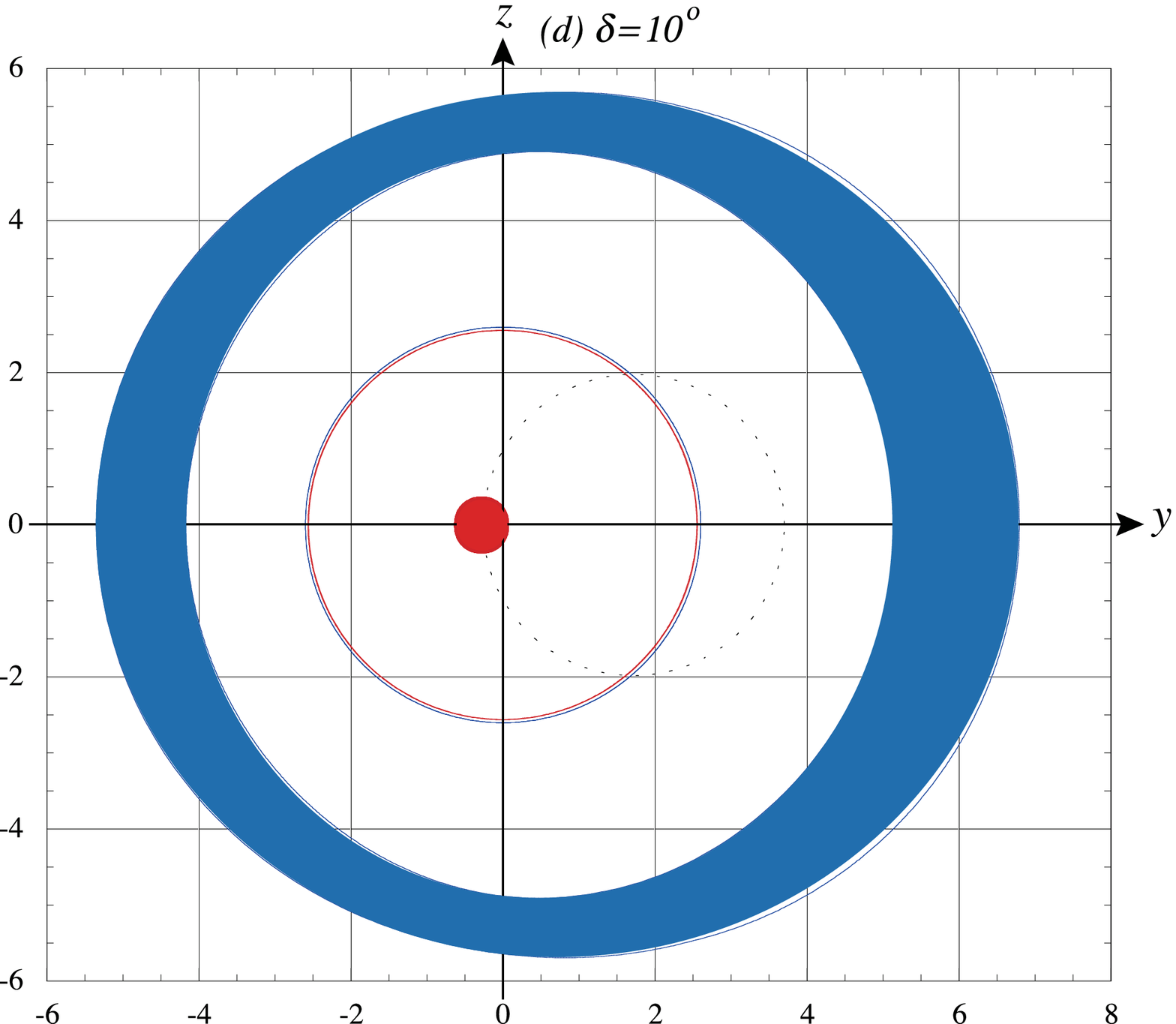}
\caption{Example of the gravitational lens images of the companion projected onto the $x=0$ plane.
We put the gravastar with $h=0.4~(\tr=1.303$) at the origin. We choose $D_s=10r_g$ and $r_s=2r_g$.
We show four snapshots when $\delta=30^\circ,~20^\circ,~12^\circ$ and $10^\circ$.
The blue images correspond to geodesics which pass only through Schwarzschild background, while the red ones to those which pass through the gravastar. The dotted lines indicate the image in the absence of the gravastar.
We do not take account of light-dark contrast, which is generated by gravitational redshift of the photons.}
\end{figure}

Next, supposing that a companion star rotates around a gravastar, we investigate its image caused by gravitational lens effects.
Figure 10 shows the setting of our numerical analysis. We suppose the companion is located behind the gravastar.
We put the gravastar's center and the companion's center at the origin and on the $z=0$-plane, respectively.
We denote the distance between the companionfs center and the gravastar's center by $D_s$ and the radius of the companion by $r_s$.
The angle $\delta$ is defined as the angle between the direction of the companion's center and the opposite direction to $x$-axis.
The image $\vec\alpha=(\alpha_y,\alpha_z)$ is defined as the intersection of the $x=0$ plane with the tangent to the ray at the observer.

Figure 11 shows an example of the images of the companion projected onto the $x=0$ plane. 
The center of the gravastar is fixed at the origin. We choose $D_s=10r_g$ and $r_s=2r_g$.
We show four snapshots when $\delta=30^\circ,~20^\circ,~12^\circ$ and $10^\circ$.
The red images correspond to geodesics which pass through the gravastar, while the blue ones to those which pass only through Schwarzschild background. The dotted circles indicate the image in the absence of the gravastar.
Here we do not take account of light-dark contrast, which is generated by gravitational redshift of the photons.
The characteristics of the gravastar are the red images: a disk in the center and arcs in the sides.
Actually, there are infinite numbers of arcs between the red arc and the blue arc.
As the companion moves, the central disk moves in the opposite direction.
As the companion approaches the center (the $x$-axis), the arcs in both sides become longer, and finally they are combined into one image, like the blue images in the Schwarzschild spacetime.

It is instructive to compare the image of the companion in Fig.\ 11 with that of the infinite plane in Fig.\ 8.
We can interpret that the red circle and the thin arcs in Fig.\ 11 are in the yellow disk and in the orange domain in Fig.\ 8, respectively.
Because the orange domain disappears when $\tr>\tr_2\approx1.58$, the blue and red thin arcs also disappear accordingly.

These lensing phenomena are interesting and give definite observational predictions.
We are afraid, however, that there may be little chance of observing images like the red ones for the following reason.
In Fig.\ 11 the red disk may be observable but are much smaller than the ``original image" represented by the dotted line.
This reduction takes place because the gravastar acts as a concave lens with much reduction rate, as discussed in Sec.\ IV.B.

\section{Concluding remarks}

Direct observation of black holes usually means observation of geometry in the vicinity of unstable circular orbits of photons.
Therefore, if there are super-compact objects which possess the unstable circular orbits, it may be difficult to distinguish them from black holes by observing photons.
As a model of super-compact objects, we have considered a model of gravastars.
We have discussed whether we can distinguish gravastars from black holes by electromagnetic observations. This work is a kind of extension of Chirenti and Rezzolla, who discussed how to tell a gravastar from a black hole by gravitational wave observations.

For definiteness, we have adopted a spherical thin-shell model of a gravastar developed by Visser and Wiltshire, which connects interior de-Sitter geometry and exterior Schwarzschild geometry.
We have found that unstable circular orbits of photons can appear around the gravastar.

Next, we have investigated the optical images of the gravastar possessing unstable circular orbits, with assuming the optically transparent surface of it and two types of optical sources behind the gravastar: (i) an infinite optical plane and (ii) a companion star.
The main feature of the image of (i) is that a bright disk and a dark thick ring surrounding the disk appear in the center of the region which would be completely dark if the compact object was not the gravastar but Schwarzschild black hole.
Also in the case (ii), a small disk and arcs around the disk appear in the region which would be completely dark for Schwarzschild black hole.

Because characteristic images appear inside the gravastar in both cases, we could tell the difference between a black hole and a gravastar with high-resolution VLBI observations near future.
It is also important to study the possibility of other quasi-black-hole objects and their observational consequences not only to discover those exotic objects but to identify black holes observationally.

Our original interest is in a gravastar which possesses unstable circular orbits of photons ($R<1.5r_g$) because the shadow of a gravastar with $R>1.5r_g$ is different from that of a black hole even if the surface is black and does not emit electromagnetic waves.
(Photons passing through the region $1.5 r_g < r < R$ make a difference in shadow between a gravastar and a black hole.)
Aside from our original interest, we have extended our analysis to the case of $R>1.5r_g$ and find that a gravastar with $1.5r_g<R<2.14r_g$ shares the above main feature: a bright disk and a dark thick ring surrounding the disk appear in the image (i).

Finally, we make a brief comment on the ``reflection" effect of a gravastar.
As shown in Figs.\ 6 and 7, there are geodesics which start at an observer, make a $\pi$ rotation about the gravastar, and then end up at the observer. That is, the observer on the Earth could detect photons which leave the Sun and travel around the gravastar.
These photons exist both in the dark domains $[a_1,a_2]$ and $[b_1,b_2]$ in Fig.\ 8.
In the dark domains photons which reach the observer come from the front side; therefore, the dark domains may be called ``mirrors" rather than ``shadows".
Such reflection phenomena have been discussed for black holes and called retro-MACHO \cite{RetroMACHO}.
What we find here is reflection phenomena also happen to photons which penetrate gravastars.

\acknowledgements
H.S. was supported by Japan Society for the Promotion of Science (JSPS), Grant-in-Aid for Scientific Research (KAKENHI, Exploratory Research, 26610050).

\end{document}